\documentclass{ws-ijmpd-mod}
\usepackage[super,compress]{cite}
\begin{document}
%------------------------------------------------------------------------------------------------------------------------------------------

\markboth{Petarpa Boonserm, Tritos Ngampitipan, and Matt Visser}
{Mimicking static anisotropic fluid spheres in general relativity}

\catchline{}{}{}{}{}
%
%------------------------------------------------------------------------------------------------------------------------------------------
%------------------------------------------------------------------------------------------------------------------------------------------
%------------------------------------------------------------------------------------------------------------------------------------------
% very standard definitions
%------------------------------------------------------------------------------------------------------------------------------------------
\def\d{{\mathrm{d}}}
%-------------------------------------------------------------------------------------------------------------------------------------------
\def\f{{\mathrm{f}}}
\def\s{{\mathrm{s}}}
\def\EM{{\mathrm{em}}}
%-------------------------------------------------------------------------------------------------------------------------------------------
\def\lint{\hbox{\Large $\displaystyle\int$}}   %needs \usepackage{amssymb} [large integral]
\def\hint{\hbox{\huge $\displaystyle\int$}}  %needs \usepackage{amssymb} [huge integral]
%------------------------------------------------------------------------------------------------------------------------------------------
%------------------------------------------------------------------------------------------------------------------------------------------
%------------------------------------------------------------------------------------------------------------------------------------------
\title{\bf\Large Mimicking static anisotropic fluid spheres in general relativity}
%------------------------------------------------------------------------------------------------------------------------------------------
\author{Petarpa Boonserm\,}
\address{Department of Mathematics and Computer Science, Faculty of Science, \\
Chulalongkorn University, Bangkok 10330, Thailand\\
petarpa.boonserm@gmail.com}
%------------------------------------------------------------------------------------------------------------------------------------------
\author{Tritos Ngampitipan\,}
\address{Department of Physics, Faculty of Science,
Chulalongkorn University, Bangkok 10330, Thailand\\
tritos.ngampitipan@gmail.com}
%------------------------------------------------------------------------------------------------------------------------------------------
\author{Matt Visser\,}
\address{ \mbox{School of Mathematics, Statistics, and Operations Research,}
Victoria University of Wellington; \\
PO Box 600, Wellington 6140, New Zealand.\\
matt.visser@msor.vuw.ac.nz}
%------------------------------------------------------------------------------------------------------------------------------------------
%------------------------------------------------------------------------------------------------------------------------------------------
\maketitle
%------------------------------------------------------------------------------------------------------------------------------------------
\begin{history}
\received{24 July 2015} %Day Month Year
\revised{7 September 2015; \LaTeX-ed \today}
\end{history}
%------------------------------------------------------------------------------------------------------------------------------------------
%------------------------------------------------------------------------------------------------------------------------------------------
\begin{abstract}
We argue that an arbitrary general relativistic static anisotropic fluid sphere, (static and spherically symmetric but with transverse pressure not equal to radial pressure), can nevertheless be successfully mimicked by suitable linear combinations of theoretically attractive and quite simple classical matter: a classical (charged) isotropic perfect fluid, a classical electromagnetic field, and a classical (minimally coupled) scalar field. While the most general decomposition is not unique, a preferred minimal decomposition can be constructed that is unique. We show how the classical energy conditions for the anisotropic fluid sphere can be related to energy conditions for the isotropic perfect fluid, electromagnetic field, and scalar field components of the model. Furthermore we show how this decomposition relates to the distribution of both electric charge density and scalar charge density throughout the model. The generalized TOV equation implies that the perfect fluid component in this model is automatically in internal equilibrium, with pressure forces, electric forces, and scalar forces balancing the gravitational pseudo-force. Consequently, we can build theoretically attractive matter models that can be used to mimic almost any static spherically symmetric spacetime.
\end{abstract}
%------------------------------------------------------------------------------------------------------------------------------------------
\keywords{Fluid spheres, anisotropic stress-energy, general relativity}
\ccode{04.20.-q 04.40.Dg 95.30.Sf}
%------------------------------------------------------------------------------------------------------------------------------------------
\maketitle
%------------------------------------------------------------------------------------------------------------------------------------------
%------------------------------------------------------------------------------------------------------------------------------------------
%------------------------------------------------------------------------------------------------------------------------------------------
%------------------------------------------------------------------------------------------------------------------------------------------
\clearpage
%%------------------------------------------------------------------------------------------------------------------------------------------
\section{Introduction}
%%------------------------------------------------------------------------------------------------------------------------------------------

Perfect fluid spheres in general relativity have almost a full century of history, dating back to Schwarzschild's interior solution of 1916~\cite{interior}, (corresponding to a spatially uniform density with a spatially varying but isotropic pressure). 
More recently, some particularly important contributions are the Delgaty--Lake review article of 1998~\cite{delgaty}, 
and the subsequent rapid (and extensive) development of ``algorithmic techniques'' in the first decade of the 21$^{st}$ century~\cite{algorithmic1, algorithmic2, algorithmic3, algorithmic4, algorithmic5, algorithmic6, algorithmic7, algorithmic8, Lake:2002, Loranger:2008, Lake:2008, Grenon:2008}.

However, when it comes to dealing with \emph{anisotropic} fluid spheres, (where the radial pressure need not equal the transverse pressure), the situation is considerably messier. See, for instance, references~\refcite{Bayin:1982}--\refcite{Schubring:2014}.
% Herrera:1997, Herrera:2001, Dehnen:2003, Herrera:2004, Herrera:2007, Viaggiu:2008, Lake:2009, Varela:2010, 
% Ivanov:2010, Ivanov:2011,  Nguyen:2013a, Nguyen:2013b, Gonzalez:2014, Schubring:2014}.  
Such anisotropic fluid spheres are sometimes believed to be physically relevant for the description of neutron stars, (see, for example, references~\refcite{Silva:2014}--\refcite{Heintzmann:1975}), and  other compact objects, and are known to be relevant for the internal structure of gravastars~\cite{gravastar1, gravastar2}.  At a pinch the formalism can also readily be adapted  to anisotropic solid spheres,  such as idealized spherically symmetric planets, or the possibly solid crust of a neutron star.
Typically  one can proceed only at the high cost of making some very specific ansatz; either for some specific metric component, or for some specific relationship between the components of the stress-energy tensor. Often these ansatze are less than well physically motivated. 

In the current article we shall endeavour instead to build a reasonably general theoretical model to ``mimic'' a generic anisotropic fluid sphere.  We shall do this by ``mimicking'' the stress-energy of a generic anisotropic fluid sphere in terms of some linear combination of  ``charged perfect fluid'' plus ``electromagnetic field'' plus ``massless scalar field''. 
By construction, the perfect fluid component in this model is automatically in internal equilibrium, with pressure (buoyancy) forces, electric forces, and scalar forces exactly balancing the gravitational pseudo-force.

We shall investigate the extent to which such a decomposition can be carried out, the extent to which such a decomposition is (or can be made to be) unique, and the extent to which we can say something concerning the spacetime geometry. 
We shall also investigate the extent to which the classical energy conditions\,~\cite{wormholes, epoch1, epoch2, epoch3, cosmo99, twilight, Bekenstein:2013} are (or can be made to be) satisfied, and the extent to which we can pin down the (effective) electric charge density and scalar charge density that are used to ``mimic'' the anisotropies in the stress-energy. 
Overall, we will seek to build a simple and straightforward theoretical model that can be usefully used to ``mimic'' a very large class of anisotropic fluid spheres. 

\enlargethispage{30pt}
For definiteness we shall adopt the conventions that the spacetime metric is presented in the usual Schwarzschild curvature coordinates
\begin{equation}
\d s^2 = - e^{-2\Phi(r)}\d t^2 + {\d r^2\over1-2m(r)/r} +r^2(\d\theta^2+\sin^2\theta\; \d\varphi^2),
\end{equation}
and that in the associated orthonormal basis the \emph{total} stress-energy is:
\begin{equation}
T^{\hat a\hat b} = 
\left[ \begin{array}{cccc} \rho & 0& 0& 0\\ 0& p_r &0 &0 \\ 0&0 & p_t &0 \\0 &0 &0 & p_t\end{array}\right].
\end{equation}

%------------------------------------------------------------------------------------------------------------------------------------
\section{Stress-energy tensors}
%-----------------------------------------------------------------------------------------------------------------------------------

\noindent
The relevant stress-energy tensors we use are utterly standard, see for instance references~\refcite{Hell} and \refcite{BD}.
\begin{itemize}
\item 
For a perfect fluid we have
\begin{equation}
T_\f^{ab} =  (\rho_\f+p_\f) V^a V^b + p_\f \, g^{ab}.
\end{equation}
\item
For the electromagnetic  field we have
\begin{equation}
T_\EM^{ab} =  F^{ac} g_{cd} F^{bd} - {1\over4} g^{ab} (F_{cd} F^{cd}).
\end{equation}
\item
For the (minimally coupled) massless scalar field we have
\begin{equation}
T_\s^{ab} =\phi^{;a} \phi^{;b} -  {1\over2} g^{ab}  (g^{cd}\phi_{;c}\phi_{;d}).
\end{equation}
\end{itemize}
Once one restricts to spherical symmetry, and adopts an orthonormal  basis, the stress-energy for a perfect fluid is:
\begin{equation}
T^{\hat a\hat b}_\f =
\left[ \begin{array}{cccc} \rho_\f & 0& 0& 0\\ 0& p_\f &0 &0 \\ 0&0 & p_\f &0 \\0 &0 &0 & p_\f\end{array}\right].
\end{equation}
Similarly, spherical symmetry plus an  orthonormal basis imply that the electromagnetic field-strength tensor is:
\begin{equation}
F^{\hat a\hat b} = E
\left[ \begin{array}{rrrr} 0 & +1& 0& 0\\  -1 &0 &0 &0 \\ 0&0 & 0 &0 \\0 &0 &0 & 0\end{array}\right].
\end{equation}
Consequently, spherical symmetry, an orthonormal basis, and tracelessness implies that the electromagnetic 
stress-energy is:
\begin{equation}
T^{\hat a\hat b}_\EM= {1\over2} E^2 
\left[ \begin{array}{rrrr} +1 & 0& 0& 0\\ 0& -1 &0 &0 \\ 0&0 & +1 &0 \\0 &0 &0 & +1\end{array}\right].
\end{equation}
Finally, using spherical symmetry and an orthonormal basis, for a scalar field:
\begin{equation}
\phi^{;\hat a} \phi^{;\hat b} = (\nabla \phi)^2 
\left[ \begin{array}{rrrr} 0 & 0& 0& 0\\ 0 &+1 &0 &0 \\ 0&0 & 0 &0 \\0 &0 &0 & 0\end{array}\right].
\end{equation}
Consequently, the stress-energy for a massless minimally coupled scalar is:
\begin{equation}
T^{ab}_\s = {1\over2} (\nabla \phi)^2 
\left[ \begin{array}{rrrr} +1 & 0& 0& 0\\ 0& +1 &0 &0 \\ 0&0 & -1 &0 \\0 &0 &0 & -1\end{array}\right].
\end{equation}

Note that for the electromagnetic field we automatically have $(p_r)_\EM <  (p_t)_\EM$, while in contrast for the scalar field  $(p_r)_\s>  (p_t)_\s$.
It is this relative magnitude flip that is essential in mimicking an arbitrary anisotropic fluid sphere.

\bigskip
\noindent
Combining these results we now have:
\begin{eqnarray}
\rho &=&  \rho_\f +{1\over2} E^2 + {1\over2} (\nabla\phi)^2 ;\\
p_r &=&  p_\f -{1\over2} E^2 + {1\over2} (\nabla\phi)^2 ;\\
p_t &=&  p_\f +{1\over2} E^2 - {1\over2} (\nabla\phi)^2 .
\end{eqnarray}
Then in particular
\begin{equation}
p_r+p_t = 2 p_\f; \quad\hbox{and} \quad  p_r-p_t = (\nabla\phi)^2-E^2.
\end{equation}
One specific way of inverting this, (by far the simplest),  is to take:
\begin{eqnarray}
p_\f &=& {1\over2}(p_r+p_t); 
\\
 \rho_\f &= &\rho - {1\over2}\,|p_r-p_t|;
 \\
 (\nabla \phi)^2 &=& \max\{p_r-p_t,0\}; 
 \\
 E^2 &=& \max\{p_t-p_r,0\}.
\end{eqnarray}
By using this decomposition we see that \emph{any} static and spherically symmetric stress tensor can be mimicked by a linear superposition of 
\begin{equation*}
\hbox{(perfect fluid)} + \hbox{(electromagnetic)} + \hbox{(massless scalar)}.
\end{equation*}
Indeed:
\begin{itemize}
\item
If $p_r\leq p_t$  everywhere within the anisotropic fluid sphere, then one can get away with only using  (perfect fluid) +  (electromagnetic)  contributions. 
\item 
If $p_r\geq p_t$  everywhere within the anisotropic fluid sphere, then one can get away with only using  (perfect fluid) + (massless scalar) contributions. 
\end{itemize}
Note that in regions where the transverse pressure is greater than the radial pressure the model has $|\nabla\phi|=0$, so the scalar field is constant. 
(We shall soon see that the scalar charge density is zero in such regions.) 
In such regions $E(r) \neq 0$, and the net charge inside the sphere of radial coordinate $r$ is simply $Q(r) = 4\pi \,E(r) \, r^2$. 
Unless $E(r)$ is fine-tuned so that $E(r)\propto 1/r^2$, there will be a non-zero electric charge \emph{density} in such regions.  
The electrically charged perfect fluid is then subject to both electric and pressure (buoyancy) forces, as well as the gravitational pseudo-force. 
Because the model is (by construction) static, these forces will automatically balance each other so that the model is in internal equilibrium --- more details on this point below, when we discuss the generalized TOV equations.  

\clearpage
If the surface of the anisotropic  fluid  occurs in such a region, where the transverse pressure is greater than the radial pressure, then in order to match an exterior Schwarzschild geometry the electric field $E(r)$ would have to be discontinuous at the surface.  (That is, $E(r_s^-)\neq 0$, but $E(r_s^+)=0$.) 
This implies the need for a surface charge density to cancel the immediate sub-surface electric field.
This is merely a slightly unusual feature of our model, it is in no sense problematic. 

If one violently objects to the presence of surface charge density, then one could instead simply match the anisotropic sphere to a Reissner--Nordstr\"om spacetime; however that would leave the anisotropic fluid sphere with a net electric charge, which we would expect to be rapidly neutralised by quite standard astrophysical processes. Overall, it seems more appropriate to work with a non-zero surface charge density but net overall charge of zero.

\bigskip
In counterpoint, in those regions where the radial pressure is greater than the transverse pressure our model has $E=0$; this corresponds to both zero electric charge density \emph{and} the somewhat stronger requirement that the \emph{net} electric charge (interior to this region) be zero. Since $\nabla\phi \neq 0$ in such regions there will now typically be a non-zero scalar charge density. 
The scalar-charged perfect fluid is then subject to both scalar forces and pressure (buoyancy) forces, as well as the gravitational pseudo-force. Because the model is (by construction) static, these forces will automatically balance each other --- more details on this point below.  If the surface of the anisotropic  fluid  occurs in such a region then to match an exterior Schwarzschild geometry the scalar field $\phi(r)$ must be continuous at the surface, and constant outside the anisotropic fluid sphere.  The matching to an exterior Schwarzschild geometry is in this case  trivial.

%------------------------------------------------------------------------------------------------------------------------------------
\section{Stress-energy tensors and their covariant divergences}
%-----------------------------------------------------------------------------------------------------------------------------------

We now proceed to set up a generalized TOV system of equations, ultimately based, (as usual), on the covariant conservation $T^{ab}{}_{;b}=0$ of the \emph{total} stress-energy. (This allows us to study the internal forces acting on the various matter components in our model, and so directly address questions of internal equilibrium.) To do so, let us first calculate the covariant divergence for each component separately. 

\medskip
\noindent
For a perfect fluid it is a standard result that\,~\cite{Hell, BD, MTW}
\begin{eqnarray}
T_\f^{ab}{}_{;b} &=& \left\{ (\rho_\f+p_\f) V^a{}_{;b} V^b + (g^{ab} + V^a V^b) (p_\f)_{;b} \right\} 
%\nonumber\\
%&&
+ \left\{ (\rho V^b)_{;b} + p (V^b{}_{;b}) \right\} V^a.\;
\end{eqnarray}
Here the two terms are by construction 4-orthogonal. One of these terms relates the 4-acceleration (in the present context, the gravitational pseudo-force) to the pressure gradient (buoyancy forces), the other ultimately provides a conservation law.

\medskip
\noindent
For the electro-magnetic field one has
\begin{equation}
T_{\EM}^{ab}{}_{;b}=  F^{ac} g_{cd} (F^{bd}{}_{;b}) +   g^{am} F_{md;b} F^{bd}  -{1\over2} g^{ab} F_{cd;b} F^{cd}.
\end{equation}
Use of the Bianchi identity for $F$, and a few standard manipulations, quickly reduces this to\,~\cite{Hell, BD, MTW} 
\begin{equation}
T_{\EM}^{ab}{}_{;b}=  - F^{ac} g_{cd} (F^{db}{}_{;b}).
\end{equation}
Now writing this in terms of the 4-current
\begin{equation}
(F^{db}{}_{;b})  = J^d = \sigma_{\EM} V^d,
\end{equation}
where $\sigma_{em}$ is the electric charge density, we have the standard Lorentz force law\,~\cite{Hell, BD, MTW}
\begin{equation}
T_{\EM}^{ab}{}_{;b}=  - F^{ab} (\sigma_{\EM} V_b) =  - \sigma_{\EM} \; F^{ab}  V_b.
\end{equation}
Note that, simply because we have assumed everything is static, the 4-current is automatically parallel to the 4-velocity of the perfect fluid component.

\medskip
\noindent
For the (minimally coupled) massless scalar field a brief computation yields\,~\cite{Hell, BD, MTW}
\begin{equation}
T_\s^{ab}{}_{;b} = \phi^{;a}{}_{;b} \phi^{;b}  + \phi^{;a}  \phi^{;b}{}_{;b}  - g^{ab}  (g^{cd} \phi_{;cb}\phi_d)
=  \phi^{;a}  (\phi^{;b}{}_{;b}).
\end{equation}
Now set $ (\phi^{;b}{}_{;b}) = \sigma_\s$, where $\sigma_\s$ is the scalar charge density.  Then the scalar force density is given by
\begin{equation}
T_\s^{ab}{}_{;b}  = \sigma_s\; \phi^{;a}.
\end{equation}

\bigskip
\noindent
Finally, combine the three (perfect~fluid), (electro-magnetic~field), and (scalar field) components. \emph{In the absence of other forms of matter}, because the \emph{total} stress-energy must be covariantly conserved, $T^{ab}{}_{;b}=0$, we have
\begin{eqnarray}
&&\left\{ (\rho_\f+p_\f) V^a{}_{;b} V^b + (g^{ab} + V^a V^b) (p_\f)_{;b} \right\} 
%\nonumber\\
%&&
%\qquad 
+ \left\{ (\rho_\f V^b)_{;b} + p_\f (V^b{}_{;b}) \right\} V^a
\nonumber\\
&&
\qquad - F^{ab} (\sigma_{\EM} V_b) +  \sigma_{\s} \phi^{;a} =0.
\end{eqnarray}
Projecting this first along the fluid 4-velocity $V$, and then perpendicular to $V$, we have the two equations
\begin{equation}
 (\rho_\f V^b)_{;b} + p_\f (V^b{}_{;b}) =  - \sigma_\s \phi_{;a} V^a,
 \label{E:1}
\end{equation}
and
\begin{eqnarray}
&& (\rho_\f+p_\f) V^a{}_{;b} V^b + (g^{ab} + V^a V^b) ([p_\f]_{;b} + \sigma_\s \phi_{;b} ) 
%\nonumber\\
%&&\qquad 
- F^{ab} (\sigma_{\EM} V_b) =0.
\label{E:2}
\end{eqnarray}
These two equations nicely summarize how the scalar charge and electric charge interact with the perfect fluid pressure, perfect fluid density, and 4-acceleration.
In particular, recall that the spacetime (and the matter content) has been assumed static, (so the timelike Killing vector $K^a$ is parallel to the fluid 4-velocity $V^a$). Then things simplify nicely.
First of all,  $\phi_{;a} V^a=0$, but we also have both $\rho_{;a} V^a=0$ and $p_{;a} V^a=0$. 
This then leads to $V^a{}_{;a}=0$, and in fact the first equation (\ref{E:1}) is vacuous.  
%\medskip
However the second equation (\ref{E:2}) is  very definitely non-vacuous ---  it simplifies only slightly to yield 
\begin{equation}
 (\rho_\f+p_\f) V^a{}_{;b} V^b + g^{ab} ([p_\f]_{;b} + \sigma_\s \phi_{;b} )  - F^{ab} (\sigma_{\EM} V_b) =0.
\label{E:2b}
\end{equation}
This will quickly lead to the generalized TOV equations. 

%-----------------------------------------------------------------------------------------------------------------------------------
\section{Generalized TOV system of equations}
%-----------------------------------------------------------------------------------------------------------------------------------

Now consider again equation (\ref{E:2b}), and note that due to spherical symmetry the three individual pieces of this equation can point only in the radial direction. Let 
\begin{equation}
n^a = \left(0,\sqrt{1-2m(r)/r},0,0\right)
\end{equation}
be the unit vector in the radial direction. Then from equation (\ref{E:2b}) we have
\begin{equation}
-  (\rho_\f+p_\f) V^a V^b n_{a;b} + n^a  ([p_\f]_{;b} + \sigma_\s \phi_{;b} ) + \sigma_\EM E = 0.
\end{equation}
But we know that if $\sigma_\s=0=\sigma_{\EM}$ this must be equivalent to the usual TOV system of equations
\begin{equation}
{\d p_\f\over\d r} = -{(\rho_\f+p_\f)(m+4\pi\, p_\f \,r^3)\over r^2(1-2m/r)};
\qquad
{\d m\over\d r} = 4\pi \rho_p r^2.
\end{equation}
Consequently, reinstating  the charge densities by taking $\sigma_\s\neq0\neq\sigma_{\EM}$, we see 
\begin{equation}
{\d p_\f\over\d r}+ \sigma_\s{\d \phi\over\d r} = -{(\rho_\f+p_\f)(m+4\pi \,p_\f \,r^3)\over r^2(1-2m/r)} - {\sigma_{\EM} E\over\sqrt{1-2m/r}}.
\end{equation}
That is, the generalized TOV equation is now
\begin{equation}
{\d p_\f\over\d r} = -{(\rho_\f+p_\f)(m+4\pi p_\f r^3)\over r^2(1-2m/r)} 
- {\sigma_{\EM} E\over\sqrt{1-2m/r}}
-  \sigma_\s {\d \phi\over\d r},
\end{equation}
supplemented by the generalized mass relation
\begin{equation}
{\d m\over\d r} = 4\pi \rho \, r^2 = 4\pi (\rho_\f +\rho_\EM+\rho_s)  r^2.
\end{equation}
Note that the first of these two equations explicitly and automatically balances the pressure force, electric force, and scalar force against the pseudo-force due to gravity. (This is as it must be, since we started from a static model there can be no motion of the perfect fluid component.)

\bigskip
\noindent
Since this is a tricky point let us be explicit: Simply by construction the model is automatically in internal equilibrium. 
We cannot however, as yet say whether this equilibrium is stable, neutral, or unstable --- at least not without additional hypotheses,  (such as some equation of state relating pressure, the mass density, and the charge densities). 
 In general, the perfect fluid need not, of course, be homogeneous, but this was already true in the Schwarzschild interior geometry of 1916, and so should be no particular surprise in this more general context. 
Similarly, while the Schwarzschild interior solution is by construction automatically in internal equilibrium, we cannot say whether this equilibrium is stable, neutral, or unstable without some additional hypothesis. 
In this regard the model we are adumbrating  in the current article is certainly no worse than the usual models for perfect fluid spheres.

\medskip
\noindent
This generalized system of TOV equations still has to be supplemented with the equations of motion for the electromagnetic and scalar fields. 
Specifically, for the electric field we have (in the quite usual fashion)
\begin{equation}
E(r) = {Q(r)\over4\pi r^2}.
\end{equation}
Here $Q(r)$ is the total net charge interior to radius $r$. Explicitly 
\begin{equation}
E(r) = {1\over4\pi r^2} \int_0^r {4\pi \bar r^2\over\sqrt{1-2m(\bar r)/\bar r}} \; \sigma_{\EM}(\bar r) \; \d \bar r.
\end{equation}
For the scalar field we have
$\Delta \phi = \sigma_\s$, which reduces to
\begin{equation}
  {1\over\sqrt{-g}} \partial_r \Big( {\sqrt{-g}\; (1-2m/r)} \partial_r\phi\Big) = \sigma_\s(r).
\end{equation}
That is
\begin{equation}
\partial_r \left( e^\Phi r^2\sqrt{1-2m/r} \; \partial_r\phi\right) = \sigma_\s(r) \, e^\Phi r^2.
\end{equation}
This can now be integrated to explicitly yield
\begin{eqnarray}
\nabla_{\hat r} \phi &=& \sqrt{1-2m/r} \; \partial_r\phi 
%\nonumber\\
%&=& 
=
{1\over e^{\Phi(r)} r^2} \int_0^r \sigma_\s(\bar r)  \, e^{\Phi(\bar r)} {\bar r}^2 \, \d \bar r.
\end{eqnarray}
If one desires an explicit formula for the scalar field itself then, up to an arbitrary additive constant, we have
\begin{eqnarray}
&&\phi(r) = \phi(0) 
%\\
%&&
+  \lint_{\!\!\!\!\!0}^r {1\over \sqrt{1-{2m(\tilde r)\over{\tilde r}}} \, e^{\Phi(\tilde r)} \tilde r^2} \left\{\int_0^{\tilde r} \sigma_\s(\bar r)  \, e^{\Phi(\bar r)} {\bar r}^2 \, \d \bar r\right\} \d \tilde r.
%\nonumber
\end{eqnarray}
This is now a complete set of differential and integral equations which uniquely determine the three-component model used to mimic the anisotropic fluid sphere. 

%------------------------------------------------------------------------------------------------------------------------------------
\section{Classical energy conditions}
%------------------------------------------------------------------------------------------------------------------------------------
Let us now consider the classical energy conditions, see for instance references \refcite{wormholes}--\refcite{Bekenstein:2013}.
%~\refcite{wormholes,epoch1,epoch2,epoch3,cosmo99,twilight,Bekenstein:2013}. 
Since  the electromagnetic and scalar field stress energy tensors both satisfy all the classical energy conditions, it is clear that provided the fluid component satisfies the classical energy conditions, then the total stress-energy must also satisfy the energy conditions. (The converse need not always hold.) 

Specifically, let us consider the NEC and note that
\begin{equation}
\rho_\f +p_\f = \rho  +  {p_r+p_t\over2} - {1\over2}\,|p_r-p_t|.
\end{equation}
This implies either
\begin{equation}
p_r > p_t: \qquad\qquad \rho_\f +p_\f = \rho  +  p_t;
\end{equation}
or
\begin{equation}
p_r < p_t: \qquad\qquad \rho_\f +p_\f = \rho  +  p_r.
\end{equation}
Therefore
\begin{equation}
\rho_\f +p_\f = \rho  +  \min\{p_r,p_t\}.
\end{equation}
That is, if the the total stress-energy satisfies the NEC, then so does the perfect fluid portion.
(Likewise, if the the total stress-energy violates the NEC then so does the perfect fluid portion.)

Similarly, consider the DEC (dominant energy condition).
The DEC requires (among other things) $\rho\pm p\geq 0$. Now 
\begin{eqnarray}
\rho_\f-p_\f &=& \rho - {(p_r+p_t)\over2} - {1\over2}\,|p_r-p_t| 
%\nonumber\\
%&=&
=
 \rho - \max\{p_r,p_t\};  
\end{eqnarray}
Combining this with the previous result for the NEC, we see that if the DEC is satisfied for the total matter, then the DEC is satisfied for the fluid component.

In contrast, consider the TEC (trace energy condition, $\rho-3p\geq 0$). Note the TEC is not fundamental physics, and was for many years abandoned~\cite{twilight}. But the TEC has undergone a recent resurgence of interest, due mainly to the fact that it can nevertheless usefully be used to characterize normal laboratory matter~\cite{Bekenstein:2013}. We note
\begin{equation}
\rho_\f-3p_\f = \rho - {3(p_r+p_t)\over2} - {1\over2}\,|p_r-p_t|.
\end{equation}
Then
\begin{eqnarray}
\rho_\f-3p_\f  &=& \rho - {(p_r+2p_t)} +p_t - \max\{p_r,p_t\}  
%\nonumber\\
% &\leq&
 \leq \rho - {(p_r+2p_t)} . 
\end{eqnarray}
So if the fluid component satisfies the TEC then the total matter satisfies the TEC.  (In this particular case, the converse need not hold.) 

In summary, we see that the various classical energy conditions can quite easily be investigated in terms of the three-component model that we have set up to mimic any anisotropic fluid sphere.  If there are any violations of the classical energy conditions,  they are confined to the perfect fluid component of the model.

%%-----------------------------------------------------------------------------------------------------------------------------------
\section{Discussion}
%%-----------------------------------------------------------------------------------------------------------------------------------

So what have we accomplished here? On the one hand, we have constructed an explicit  three-component model that is capable of mimicking any anisotropic fluid sphere, with two of the components (electro-magnetic field and scalar field)  automatically being theoretically attractive and physically well behaved (eg, satisfying the classical energy conditions). 
It is only the perfect fluid component that has any risk of being ``exotic''  (ie, violating one or more of the classical energy conditions). 

On the one hand this is a very powerful result; on the other hand it is perhaps a little too powerful. Note that we are not claiming that any anisotropic fluid sphere \emph{is} physically one of these three-component models, but are instead making the more modest claim that both the spacetime geometry and the total stress-energy can be successfully mimicked by one of these three-component models. 

We feel that this is already an interesting observation, and that these three-component models may be of some theoretical (and maybe even practical) interest --- certainly, since our construction is entirely generic, one can eliminate the need for making any \emph{ad hoc} choices for the metric components or the equation of state, as is typically done for currently extant models of anisotropic fluid spheres. 

%\bigskip

%\newpage
%------------------------------------------------------------------------------------------------------------------------------------------
\section*{Acknowledgments}
%------------------------------------------------------------------------------------------------------------------------------------------

This research has been supported by Ratchadapisek Sompoch Endowment Fund, Chulalongkorn University (Sci-Super 2014-032), and by a grant for the professional development of new academic staff from the Ratchadapisek Somphot Fund at Chulalongkorn University, by the Thailand Toray Science Foundation (TTSF), by the Thailand Research Fund (TRF), by the Office of the Higher Education Commission (OHEC), Faculty of Science, Chulalongkorn University (MRG5680171), and by the Research Strategic plan program (A1B1), Faculty of Science, Chulalongkorn University.
PB was additionally supported by a scholarship from the Royal Government of Thailand. TN was also supported by a scholarship from the Development and Promotion of Science and Technology talent project (DPST).

MV was supported by the Marsden Fund, via a grant administered by the Royal Society of New Zealand.

%------------------------------------------------------------------------------------------------------------------------------------------
\appendix
%------------------------------------------------------------------------------------------------------------------------------------------
\section{General solution to the inversion problem}
%------------------------------------------------------------------------------------------------------------------------------------------

\noindent
The \emph{completely general} solution to the inversion problem, now parameterized by an arbitrary function $h(r)$,  is
\begin{eqnarray}
p_\f &=& {p_r+p_t\over2};  
\\
 (\nabla \phi)^2 &=& \max\{p_r-p_t,0\} + h^2; 
 \\
E^2 &=& \max\{p_t-p_r,0\} + h^2;
\\
\rho_\f &=& \rho - {1\over2}\,|p_r-p_t| - h^2. 
\end{eqnarray}
Note that at the surface of the anisotropic fluid sphere we now have $|E(r_s^-)| \geq |h(r_s^-)|$. So if $|h(r_s^-)|>0$, and if we are to embed this model into an exterior Schwarzschild geometry, then there must be a compensating shell of electric charge density at the surface. 
This is merely a feature of the model, it is in no sense problematic. 

With regards to the NEC, note that in particular
\begin{equation}
\rho_\f +p_\f = \rho  +  {p_r+p_t\over2} - {1\over2}\,|p_r-p_t| - h^2.
\end{equation}
So in this case we have either
\begin{equation}
p_r > p_t: \qquad\qquad \rho_\f +p_\f = \rho  +  p_t  - h^2;
\end{equation}
or
\begin{equation}
p_r < p_t: \qquad\qquad \rho_\f +p_\f = \rho  +  p_r  - h^2.
\end{equation}
Therefore 
\begin{equation}
\rho_\f +p_\f = \rho  +  \min\{p_r,p_t\} - h^2.
\end{equation}

\noindent
Note that statements about the NEC are now trickier:
\begin{itemize}
\item Violation of the total NEC now implies violation of the NEC for the fluid component.
\item Violation of the fluid component NEC no longer implies violation of the total NEC. 
\item Even if the total stress-energy satisfies the NEC, one can now always (by choosing $h^2$ large enough) force the fluid component to violate the NEC. (Or force $w_\f=-1$, on the border of NEC violation. Note that a fluid component satisfying $\rho_\f+p_\f=0$ is now \emph{not} cosmological constant because it is interacting with other forms of matter.)
\end{itemize}

Regarding the  DEC (the dominant energy condition), which
 requires $\rho\pm p\geq 0$, let us consider
\begin{eqnarray}
\rho_\f-p_\f &=& \rho - {(p_r+p_t)\over2} - {1\over2}\,|p_r-p_t| - h^2
%\nonumber\\
%&=& 
=\rho - \max\{p_r,p_t\} - h^2.
\end{eqnarray}
So if the DEC  is satisfied for the total matter, one can always choose $h$ to make the fluid  ``stiff'': $\rho_\f=p_\f$.

When it comes to the TEC (the trace energy condition, $\rho-3p\geq 0$, see references ~\refcite{twilight} and~\refcite{Bekenstein:2013}), we have
\begin{equation}
\rho_\f-3p_\f = \rho - {3(p_r+p_t)\over2} - {1\over2}\,|p_r-p_t| - h^2.
\end{equation}
Then
\begin{eqnarray}
\rho_\f-3p_\f 
&=& \rho - {(p_r+2p_t)} +p_t - \max\{p_r,p_t\} - h^2   
%\nonumber\\
%&\leq& 
\leq \rho - {(p_r+2p_t)} ;  
\end{eqnarray}
So if the perfect fluid component satisfies the TEC, then the total matter content satisfies the TEC. 

\enlargethispage{10pt}
Now consider the Strong Energy Condition (SEC), which (in addition to the NEC) states that $\rho + 3p \geq 0$. Then
\begin{eqnarray}
\rho_{\f} + 3p_{\f} &=& \rho + \frac{3(p_{r} + p_{t})}{2} - \frac{1}{2}|p_{r} - p_{t}| - h^{2}\nonumber\\
                  &=& \rho + (p_{r} + p_{t}) + \frac{p_{r} + p_{t}}{2} - \frac{1}{2}|p_{r} - p_{t}| - h^{2}\nonumber\\
                  &=& \rho + (p_{r} + p_{t}) + \min\{p_{r}, p_{t}\} - h^{2}\nonumber\\
                  &=& \rho + (p_{r} + 2p_{t}) - p_{t} + \min\{p_{r}, p_{t}\} - h^{2}\nonumber\\
                  &\leq& \rho + (p_{r} + 2p_{t}).
\end{eqnarray}
So if the fluid component satisfies the SEC, then the total matter content satisfies the SEC.\\

%\clearpage
%------------------------------------------------------------------------------------------------------------------------------------------

%------------------------------------------------------------------------------------------------------------------------------------------
%\null
%\enlargethispage{125pt}
%\vspace{-60pt}
%%------------------------------------------------------------------------------------------------------------------------------------------
%\begin{thebibliography}{69}
%%------------------------------------------------------------------------------------------------------------------------------------------
%% trick for a horizontal bar
%%------------------------------------------------------------------------------------------------------------------------------------------
%\end{thebibliography}
%%------------------------------------------------------------------------------------------------------------------------------------------

%-----------------------------------------------------------------------------------------------------
\end{document}